\documentclass{article}
\pdfoutput=1
\usepackage{fullpage}
\usepackage{url}
\usepackage{graphicx}
\makeatletter
\def\section{\@startsection {section}{1}{\z@}{-3.5ex plus -1ex minus
 -.2ex}{2.3ex plus .2ex}{\bf}}
\def\subsection{\@startsection {subsection}{1}{\z@}{-3.5ex plus -1ex minus
 -.2ex}{2.3ex plus .2ex}{\bf}}
\makeatother

\begin{document}
\title{{TREC-COVID: Constructing a Pandemic Information Retrieval Test Collection}}

\author{Ellen Voorhees\\
       National Institute of Standards and Technology\\
       \and
       Tasmeer Alam\\
       National Institute of Standards and Technology\\
        \and
        Steven Bedrick\\
        Oregon Health and Science University\\
        \and 
        Dina Demner-Fushman\\ 
        U.S. National Library of Medicine\\
        \and
        William R Hersh\\ 
        Oregon Health and Science University\\
        \and 
        Kyle Lo\\
        Allen Institute for AI\\
        \and 
        Kirk Roberts\\
        University of Texas Health Science Center at Houston\\
        \and
        Ian Soboroff\\
        National Institute of Standards and Technology\\
        \and 
        Lucy Lu Wang\\
        Allen Institute for AI\\
       \date{}}

\maketitle
\abstract{ 
TREC-COVID is a community evaluation designed to build a test collection that captures the information needs of biomedical researchers using the scientific literature during a pandemic.
One of the key characteristics of pandemic search is the accelerated
rate of change: the topics of interest evolve as the pandemic progresses and
the scientific literature in the area explodes. The COVID-19 pandemic
provides an opportunity to capture this progression as it happens. TREC-COVID, in creating a test collection around COVID-19 literature, is building infrastructure to support new research and technologies in pandemic search.
}

\section{Introduction}

The COVID-19 pandemic has caused not only a public health crisis but 
 also an information crisis. The global Internet can spread incorrect or incomplete information faster than the spread of the virus. Even the scientific community is challenged, as the platform for open science, especially preprints, allows scientific information to spread that has not been fully vetted. This results in a difficult tension between wanting to disseminate scientific information as quickly as possible while wanting it to be vetted (i.e., peer-reviewed) as well.

To allow scientists, clinicians, and policy makers to avail themselves of the knowledge contained within this new information landscape, the White House requested that
the Allen Institute for Artificial Intelligence and their collaborators aggregate
a structured dataset of coronavirus research for the global research community~\cite{Wang2020CORD19TC}. This COVID-19 Open Research Dataset, or CORD-19, contains scholarly articles about COVID-19 and the coronavirus family of viruses.
The articles are drawn from peer-reviewed literature in PubMed Central\footnote{\url{https://www.ncbi.nlm.nih.gov/pmc/}}
as well as from archival sites such as medRxiv\footnote{\url{https://www.medrxiv.org/}} and bioRxiv.\footnote{\url{https://www.biorxiv.org/}}
Importantly, CORD-19 is not a static document collection, but is regularly updated with new versions released once a week.
The different versions of CORD-19 capture the growth and change in the
COVID-19 literature over time.

The availability of CORD-19 provides the opportunity to create a test collection
that captures the information needs of biomedical researchers using the
scientific literature during a pandemic, and the TREC-COVID evaluation aims to do precisely that.
The twin goals of TREC-COVID are
\begin{itemize}
\item to evaluate search algorithms and systems for helping scientists, clinicians, policy makers, and others manage the existing and rapidly growing corpus of scientific literature related to COVID-19, and
\item to discover methods that will assist with managing scientific information in future global biomedical crises.
\end{itemize}
Based on the TREC\footnote{\url{http://trec.nist.gov}} framework, TREC-COVID is 
building a pandemic test collection through a series of rounds.
It builds on a history of biomedical TREC tracks~\cite{genomics,medicalRecords,Roberts15trec,Roberts17.trec},
and specifically on the Genomics, Clinical Decision Support, and Precision Medicine tracks that focused on retrieval from the scientific literature.
But as discussed in this article, TREC-COVID must also extend the evaluation procedures of these tracks to realize the goal of capturing the pandemic information environment. These extensions include multi-round evaluation, mid-evaluation updates to the topic and document sets, and tight deadlines for both participation and assessment.

As of this writing, the first round of TREC-COVID has just completed and the
second round has begun.
This paper provides a snapshot of TREC-COVID's process and results at this point in time.

\section{Structure of TREC-COVID}
Each round of TREC-COVID is structured as an independent community evaluation challenge much like a typical TREC track.
The systems' task is a classic ad~hoc search task using CORD-19 as the 
document set and a set of biomedical questions as the topics (i.e., statements of information need).
Systems produce a ranked list of documents per topic where each list is ordered by decreasing likelihood that the document matches the information need (this is a {\em run}).
The set of submitted runs are used to define a much smaller set of documents for human assessors to judge for relevance.
The documents, topics, and relevance judgments together form a test collection
that can be used to evaluate the relative effectiveness of competing retrieval approaches.

TREC-COVID differs from a typical TREC track in two important ways.
The most obvious difference is the compressed time schedule.
Because we want to foster research on systems that are able to pivot quickly, deadlines are very short: roughly one week between when the test topics
become available and the run submission deadline, followed by ten days for relevance assessing before the next round begins.

The second difference is that while each round is treated as an independent evaluation, the document and topic sets in each round are supersets of previous rounds' sets.  Relevance judgments from earlier rounds provide training data for later rounds (and also mean residual collection scoring must be done in later rounds as described below).  The cumulative document, topic, and relevance judgment sets will reflect the changes observed in real use of search systems during the current pandemic.

The remainder of this section describes the three components of the test collection in more detail.

\subsection{Documents}
\label{sec:documents}

TREC-COVID uses the document set provided by CORD-19 \cite{Wang2020CORD19TC}. CORD-19 consists of new publications and preprints on the subject of COVID-19, as well as relevant historical research on coronaviruses, including SARS and MERS. The April 10, 2020 release of CORD-19, which is used for the first round of TREC-COVID, includes 51K papers, with full text available for 39K. As of May 1, 2020, CORD-19 consists of 60K papers, of which full text is available for 48K. Due to the ongoing nature of the pandemic, the CORD-19 corpus is constantly evolving and updated regularly (approximately weekly) as more new or historical research becomes available. Each TREC-COVID round uses the release of the dataset that is the most up-to-date as of the start of the round.

Content in CORD-19 is derived from multiple sources, and harmonization logic is applied to identify and remove duplicate entries. Each resulting entry in CORD-19 is given a unique identifier, the CORD UID, which corresponds to the notion of a \emph{paper}. Each paper can be associated with multiple documents; for example, the main paper document plus supplementary files, or potentially, multiple versions of a paper such as its preprint and camera-ready editions. In TREC-COVID, provisions are made during assessment to adapt to this notion of a paper, by showing all associated documents to assessors at the time of judgement.

Researchers have also made significant use of preprints during this epidemic to rapidly disseminate contributions. Preprints, much more than publications, are subject to iteration, and content can change dramatically between versions. Due to all versions inherently being part of the same paper, the identifiers for these preprints persist across CORD-19 corpus releases. In cases where changes occur in the documents associated with the same paper (identified by UID) between CORD-19 corpus releases, prior judgments on the original documents may no longer hold.
A judgment in TREC-COVID is associated with a particular round to indicate the version of the document that received the judgment.

CORD-19 presents a real life document set that evolves in response to new research. Though there are difficulties in working with such a document set, it is an opportunity to study ways to judge changing relevance in a realistic environment. Further challenges and proposed solutions are detailed in Section \ref{sec:challenges}.

\subsection{Topics}
The topics used in TREC-COVID have been written by its organizers with biomedical 
training, inspired by consumer questions submitted to the National 
Library of Medicine, discussions by medical influencers on social media, and 
suggestions solicited on Twitter via the \#COVIDSearch tag in late March 2020.
They are representative of the high-level concerns related to the pandemic.
So far, the topics have not included detailed biological questions (e.g., RNA mutation rates, structure of a certain protein), though in future rounds more of these topics may be added.

Each topic is composed of three fields:
\begin{enumerate}
\item \textit{query}: a short keyword query
\item \textit{question}: a more precise natural language question
\item \textit{narrative}: a longer description that further elaborates on the question, often providing specific types of information that would fall under the topic score
\end{enumerate}
Note that while the \textit{question} is a superset of the information in the \textit{query}, the \textit{narrative} is not intended to be a superset of the information in the \textit{question}.
Instead, it helps to further specify user intent.
For three examples of topics from Round 1, see Figure~\ref{topics.fig}.

There was considerable discussion amongst the organizers prior to Round 1---and some discussion with participants after Round 1---relating to the terminology used to refer to the virus/disease.
Early real-world queries referred to the virus in informal terms (e.g., ``\textit{Wuhan virus}'', ``\textit{Chinese flu}''), then evolved to more formal terms such as ``\textit{2019-nCoV}'' and informally simply ``\textit{coronavirus}'', and finally the official use of ``\textit{SARS-CoV-2}'' to refer to the virus and ``\textit{COVID-19}'' to refer to the condition it causes.
This is a fundamental linguistic problem in dealing with an emerging situation like a pandemic.
A specific problem is that ``\textit{coronavirus}'' refers to a class of viruses, not just SARS-CoV-2.
However, the user intent of the information need is clear:
the user is focused on COVID-19/SARS-CoV-2.
A hybrid solution was taken to this problem in topic construction.
The \textit{query} field often uses less formal terms, such as ``\textit{coronavirus}'', while the \textit{narrative} field is more likely to use a formal term such as ``\textit{COVID-19}'' or ``\textit{SARS-CoV-2}''.
At least one of the unambiguous terms is guaranteed to be used in at least one of the fields for the topic, ensuring that---to a human, at least---it is clear that the topic is about the current pandemic.
Of course, some information about prior coronaviruses is still useful in the current pandemic (which is why articles on these are included in CORD-19).
The extent to which this background information is useful depends on the topic, how much context is needed, and to some extent the user's expectation about what may be available about a novel virus.
Understanding user intent, then, is left as a challenge to the participant systems.
\begin{figure}
   \begin{center}
   \begin{tabular}{|lp{5.25in}|}
   \hline
{\bf number}: & 7\\
{\bf query}: & serological tests for coronavirus\\
{\bf question}: & are there serological tests that detect antibodies to coronavirus?\\
{\bf narrative}: & Looking for assays that measure immune response to COVID-19 that will help determine past infection and subsequent possible immunity.\\
 & \\
{\bf number}: & 20\\
{\bf query}: &  coronavirus and ACE inhibitors \\
{\bf question}: & are patients taking Angiotensin-converting enzyme inhibitors (ACE) at increased risk for COVID-19?\\
{\bf narrative}: & Looking for information on interactions between coronavirus and angiotensin converting enzyme 2 (ACE2) receptors, risk for patients taking these medications, and recommendations for these patients.\\
 & \\
{\bf number}: & 29\\
{\bf query}: & coronavirus drug repurposing\\
{\bf question}:&  which SARS-CoV-2 proteins-human proteins interactions indicate potential for drug targets. Are there approved drugs that can be repurposed based on this information?\\
{\bf narrative}: & Seeking information about protein-protein interactions for any of the SARS-CoV-2 structural proteins that represent a promising therapeutic target, and the drug molecules that may inhibit the virus and the host cell receptors at entry step.\\
\hline
\end{tabular}
\end{center}
\caption{Example topics from the Round 1 test set.}
\label{topics.fig}
\end{figure}

In future rounds, new topics will be added
(5 per round are planned).
These new topics will capture some of the trends that emerge as the crisis unfolds (especially new information after Round 1).
More detailed, scientifically challenging topics will also be added to test the range of capabilities of systems.

\subsection{Relevance Judgments}

As with the other TREC biomedical tasks, the annotators making relevance assessments for TREC-COVID are individuals with clinical expertise. To date, we have been able to enlist the help of ten Oregon Health and Science University medical students whose clinical activities were displaced due to Covid-19 restrictions, as well as additional relevance assessment help from professional indexers from the National Library of Medicine. 

\begin{figure}
    \centering
    \includegraphics[width=6in]{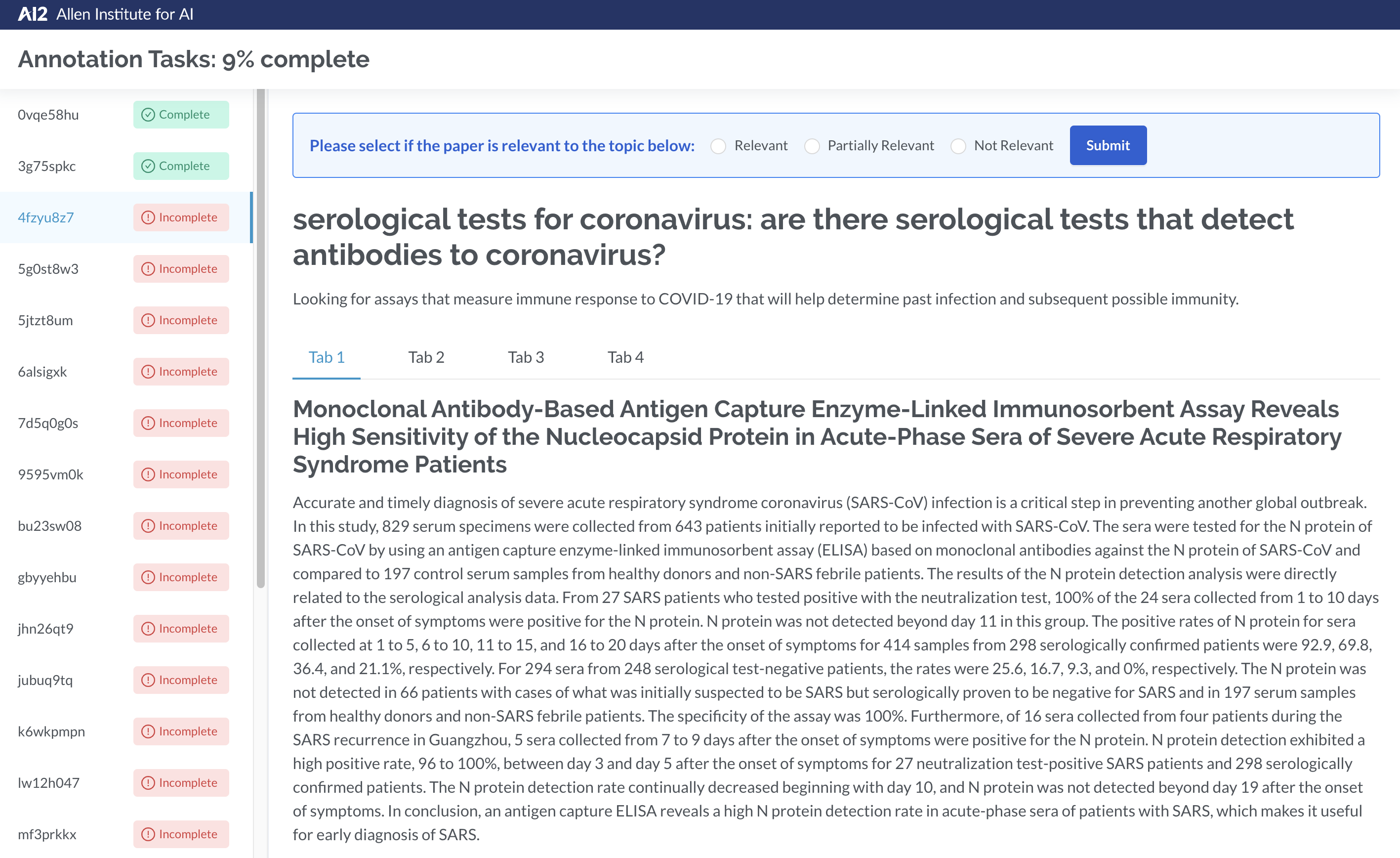}
    \caption{Screenshot of the interface used for TREC-COVID document assessing.}
    \label{assessing.fig}
\end{figure}

The relevance assessors use the Web-based system illustrated in Figure~\ref{assessing.fig}.
The assessment system shows the assessor the topic and a list of documents to be judged. Assessors mark each document in the list as either `Relevant',`Partially Relevant', or `Not Relevant'. Each document may have one or more forms, e.g., title/abstract, preprint, PDF from publisher, and/or entry into PubMed Cental that are included under different tabs. The document list on the left side of the screen also indicates whether the document has been judged.

\begin{figure}
    \centering
    \includegraphics[height=3.5in]{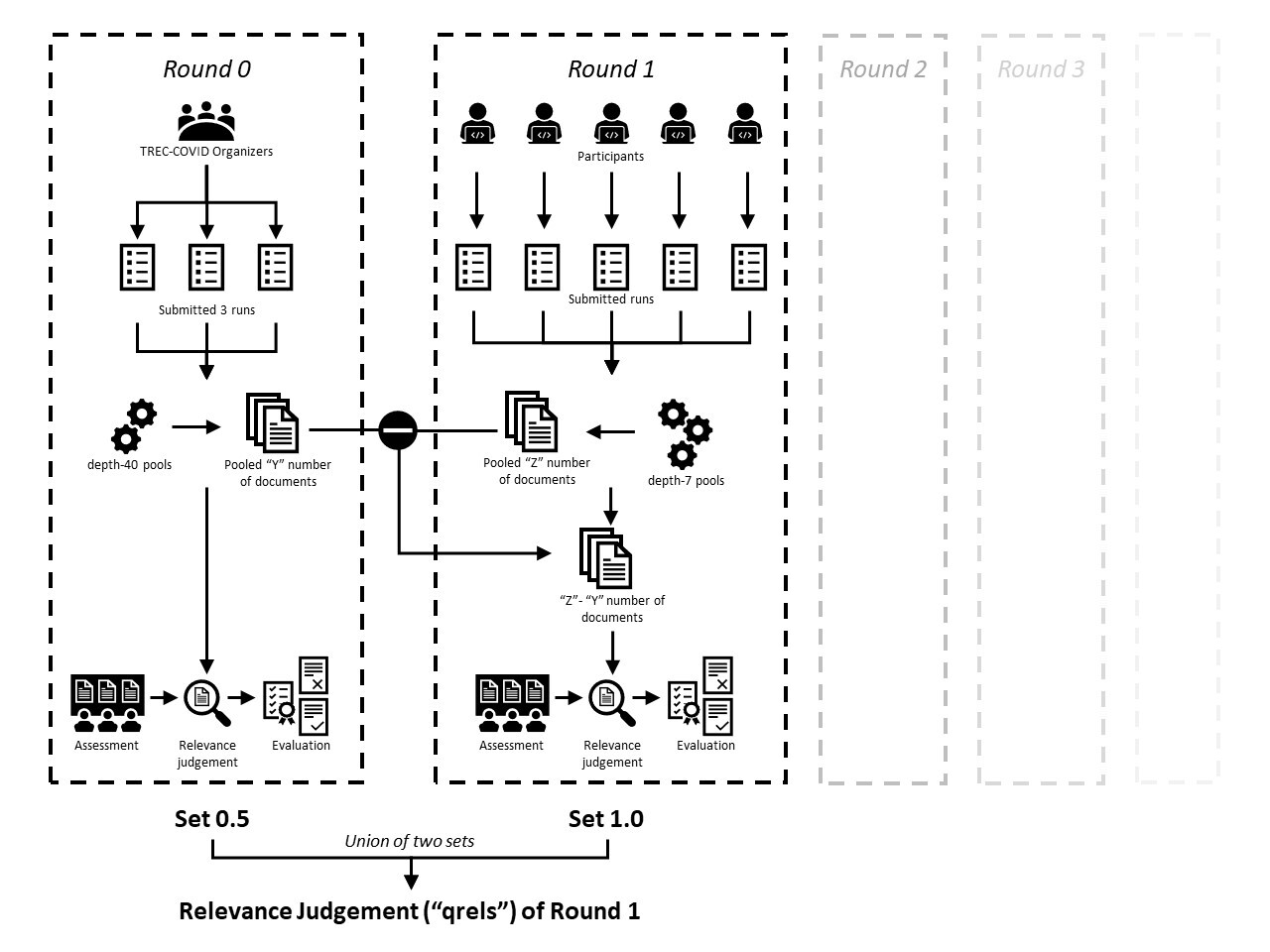}
    \caption{Illustration of the process to create the Round X qrels. The Round X qrels file is the union of the set of judgments made the week following the Round X submission deadline and the set of judgments made in the preceding week.  For the special case of ``Round 0'', organizers created three runs and pooled those to start the judgment process prior to receiving any participant runs. }
    \label{qrels.fig}
\end{figure}

As mentioned earlier, the compressed schedule of TREC-COVID means that the amount of time available for performing relevance assessments is very short.  Since computing which documents to judge is driven by the submitted runs, there are only two weeks of judging time per round. Assessments performed during the week that starts immediately after the run submission deadline of Round~X runs are called X judgments. Assessments performed in the following week when participants are constructing their Round~X+1 submissions are called X.5 judgments. 
Both judgment sets~X and~X.5 are derived from Round~X runs and use the version of CORD-19 that runs in Round~X searched.
However, the relevance judgments used to evaluate Round~X runs, and which are publicly released before Round~X+1 begins, consist of judgment sets~X-.5 and~X. 
The standard format of a relevance judgments file (the so-called {\em qrels} file) is \mbox{{\tt topic-id iteration doc-id judgment}} where the iteration field is traditionally zero.
TREC-COVID qrels files record the judgment round when the judgment was made in this field.
Note that a 0.5 judgment set exists, too. 
Because assessors were available before the first submission deadline, TREC-COVID organizers produced three runs based on the Anserini\footnote{\url{https://github.com/castorini/anserini}} system and used depth-40
pools across these three runs using the Round~1 release of CORD-19 (April~10) to create the 0.5 set.
Figure~\ref{qrels.fig} illustrates the relationship between TREC-COVID rounds and the corresponding judgment sets.

The pool of available assessors constrains the amount of judgments that can be made to about 150--200 per topic per week.
To obtain this number of assessments, a topic may be judged by more than one assessor.
For Round~1, the maximum number of assessors for a topic was two.

With only about 300 documents judged per topic in a given round, the
relevance judgments for that round are incomplete to the point that run comparisons are likely unstable for many measures.
That is, if the unjudged documents in the runs were judged, the preferred order of the two runs could change.
While full trec\_eval output as produced by the current round's qrels file is reported for each run\footnote{All submitted runs and score reports are archived at \url{http://ir.nist.gov/covidSubmit/Archive.html}.}, Round~1 summaries therefore focused on three measures: Bpref, NDCG@10, and P@5.  Bpref was designed for collections with incomplete judgments and is computed over only the judged set.
P@5 and NDCG@10 are each affected by only the most highly ranked documents, which are the documents most likely to be judged.  The computation of NDCG uses gain values
of 1 for `Partially Relevant' documents and 2 for `Relevant' documents.
Measures that use binary judgments are computed using both types of relevant documents as the relevant set.

The Round~X qrels file is posted to the TREC-COVID web site at the time the Round~X+1 test set is released.  Participants are free to use judgments from earlier rounds to construct runs for later rounds.
To properly account for the use of relevance feedback when scoring runs, each
round after the first will use {\em residual collection} evaluation~\cite{Salton1997ImprovingRP}.
In residual collection evaluation, any document that has already been judged for a topic (in any round) is removed from the collection before scoring. This is implemented by removing all documents that were judged for a topic in an earlier round from all submitted runs.

While residual collection evaluation is methodologically valid, it means that each
round's runs can only be evaluated using a single round's qrels, not the cumulative set of judgments for a topic.
Thus each round will most likely continue to be affected by incomplete judgments issues.
Further, residual collection evaluation will make effectiveness scores appear to degrade from round to round since the set of relevant documents in the collection is reduced in each round.
The final pandemic test collection that is the end product of TREC-COVID will contain the cumulative judgments from all rounds and can therefore be expected to be much more stable than the individual-round collections.

\section{Round 1 Results}

TREC-COVID Round~1 received 143 runs from 56 teams.
Teams came from Asia, Australia, Europe, and North America.
One hundred of the runs are automatic runs (a run in which the system is given the topic file and produces ranked output with no human involvement at all) and the
remaining 43 runs are manual runs (everything else).

The number and diversity of the submitted runs meant that staying within the judgment budget of 200 average documents per topic was challenging.
Anticipating the possibility that we would not be able to judge all runs,
participants were asked to assign a judging priority (1--3) to their runs at
submission time.
Depth-7 pools\footnote{See summary papers such as~\cite{philosophy} or TREC Overview papers for details on pooling.} created from only the first priority run from each
team amounted to almost 7000 documents to be judged across the 30 topics.
Subtracting the 0.5~judgments from that set left a total of just over
6000 documents.  These 6000 documents are the 1.0~judgment set.
The union of judgment sets 0.5 and 1.0, which constitute the Round 1 qrels, contains 8,691 judgments, a mean of 289.7 judgments per topic with a range between 180--373 per topic (see Table~\ref{relcounts.tab}).
\begin{table}
\caption{Counts of total numbers of judged documents and number of relevant documents per topic. Percent relevant is the fraction of judged documents that are some form of relevant.}
\label{relcounts.tab}
\begin{center}
\begin{tabular}{rrrrr|rrrrr}
 & \multicolumn{1}{c}{Total} & \multicolumn{1}{c}{Partially} & &\multicolumn{1}{c|}{Percent} &
 & \multicolumn{1}{c}{Total} & \multicolumn{1}{c}{Partially} & &\multicolumn{1}{c}{Percent}\\
\multicolumn{1}{c}{Topic} & \multicolumn{1}{c}{Judged} & \multicolumn{1}{c}{Relevant} & \multicolumn{1}{c}{Relevant}& \multicolumn{1}{c|}{Relevant} & \multicolumn{1}{c}{Topic} & \multicolumn{1}{c}{Judged} & \multicolumn{1}{c}{Relevant} & \multicolumn{1}{c}{Relevant} & \multicolumn{1}{c}{Relevant}\\ 
\hline
1 &  323 & 45 & 56 & 0.313 & 16 &  340 & 42 & 11 & 0.156\\
2 &  284 & 21 & 26 & 0.165 & 17 &  243 & 32 & 45 & 0.317\\
3 &  337 & 66 & 24 & 0.267 & 18 &  267 & 79 & 32 & 0.416\\
4 &  357 & 32 & 27 & 0.165 & 19 &  301 & 27 & 16 & 0.143\\
5 &  336 & 35 & 96 & 0.390 & 20 &  247 & 41 & 25 & 0.267\\
6 &  321 & 80 & 83 & 0.508 & 21 &  319 & 15 & 70 & 0.266\\
7 &  275 & 2 & 47  & 0.178 & 22 &  259 & 17 & 30 & 0.181\\
8 &  360 & 46 & 30 & 0.211 & 23 &  256 &  4 & 22 & 0.102\\
9 &  298 & 25 & 16 & 0.138 & 24 &  249 & 14 & 19 & 0.133\\
10 &  191 & 35 & 50 & 0.445 & 25 & 308 & 9 & 62 & 0.231\\
11 &  344 & 67 & 5  & 0.209 & 26 & 312 & 19 & 106 & 0.401\\
12 &  324 & 76 & 126 & 0.623 & 27 & 300 & 30 & 44 & 0.247 \\
13 &  373 & 97 & 49 & 0.391  & 28 & 180 & 9 & 29 & 0.211\\
14 &  222 & 24 & 5 & 0.131   & 29 & 218 & 42 & 58 & 0.459\\
15 &  348 & 45 & 12 & 0.164  & 30 & 199 & 39 & 16 & 0.276\\
\end{tabular}
\end{center}
\end{table}

\begin{figure}
    \centering
    \includegraphics[width=6.5in]{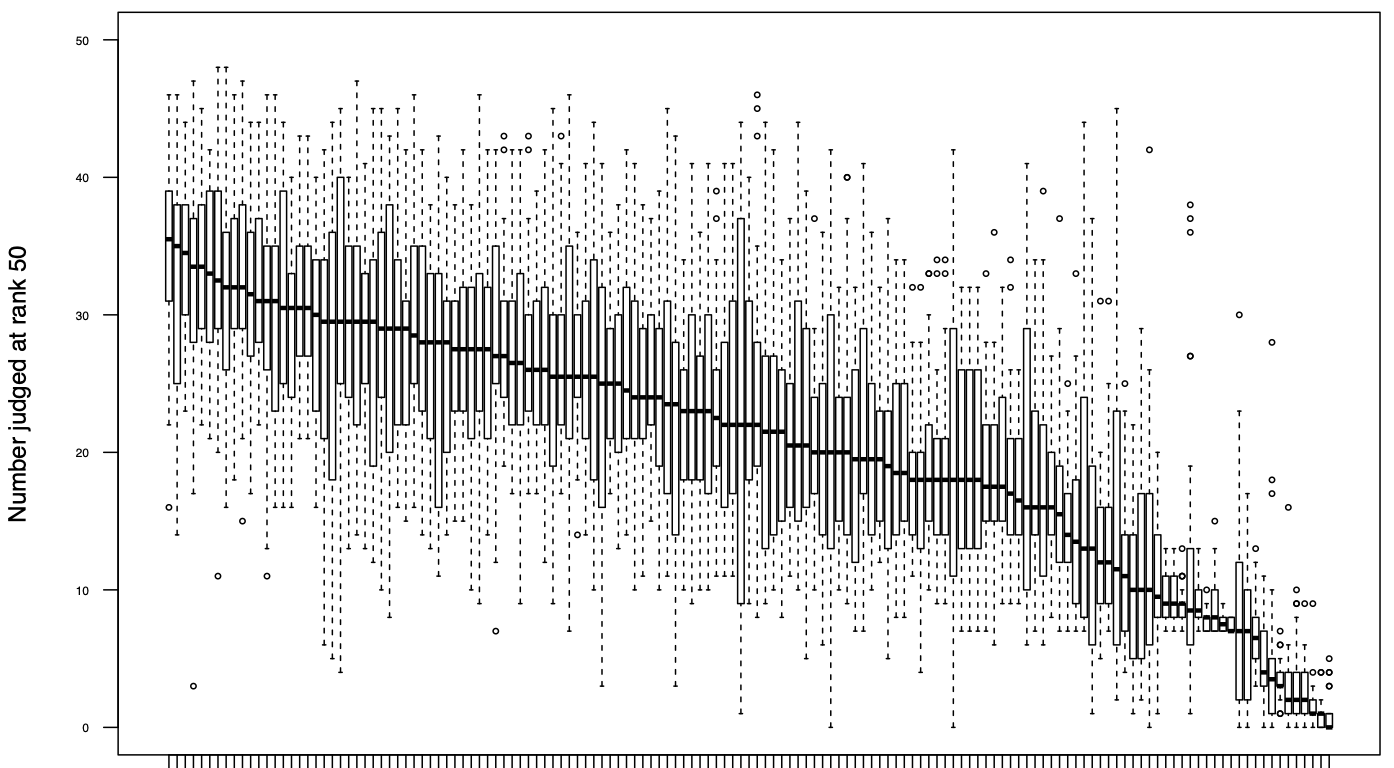}
    \caption{Number of documents in top 50 ranks that are judged over all 30 topics.}
    \label{judged.fig}
\end{figure}

The maximum size of depth-7 pools across 56 runs (one run per team) is 11,760 ($56 \times 30 \times 7$).
The 7000 size obtained in Round~1 pooling is greater than half of that maximum, which is a high percentage compared to typical TREC collections.
By definition, this means there is comparatively little overlap among top-ranked documents across runs.
Many runs do contain other runs' top documents deeper in their ranked lists, however.
Figure~\ref{judged.fig} shows a box-and-whiskers plot of the number of judged documents in the top 50 ranks per topic for all submitted runs.
The runs are plotted on the x-axis and are ordered by decreasing median number of documents judged (the heavy black line).
About half the runs have a median of approximately 25, so for those runs at least half of the topics have at least half of the top-50 documents judged.
The variance across topics is large for most runs, and there is 
a set of runs to the far right in the graph that retrieved practically no documents in common with other runs' top documents down to depth 50.

Figure~\ref{judged.fig} shows statistics computed over judged documents, not relevant documents only.
Historically, runs with extremely low overlap with other runs tend to be
runs created with errors where the assumption that unjudged documents are
not relevant holds.
We cannot know this to be true for the runs on the far right of the graph without more judgments, however.
Judgments from set 1.5, which are currently in process, may provide
a partial answer.

Table~\ref{relcounts.tab} shows relevance counts for each topic.
The table includes the total number of judged documents, the number of documents judged `Partially Relevant' and `Relevant', and the percentage of judged documents that are (some form of) relevant per topic in the Round~1 qrels.
The total number relevant per topic ranges from 26--202.
The percentage of judged that are relevant is of interest as an indicator of the stability (or reusability) of the collection.
Historically, when more than a third of the judged documents are relevant for a topic, it is highly likely that many more relevant documents that have not yet been identified remain in the collection~\cite{CIKMbandits}.  Having fewer than one third relevant is not a guarantee that the collection is stable, but more than a third is strong evidence that it is not.
For the Round~1 qrels, more than a quarter of the topics (8/30) have relevant percentages greater than 0.33.

\begin{figure}
\begin{center}
\includegraphics[height=4in,width=5in]{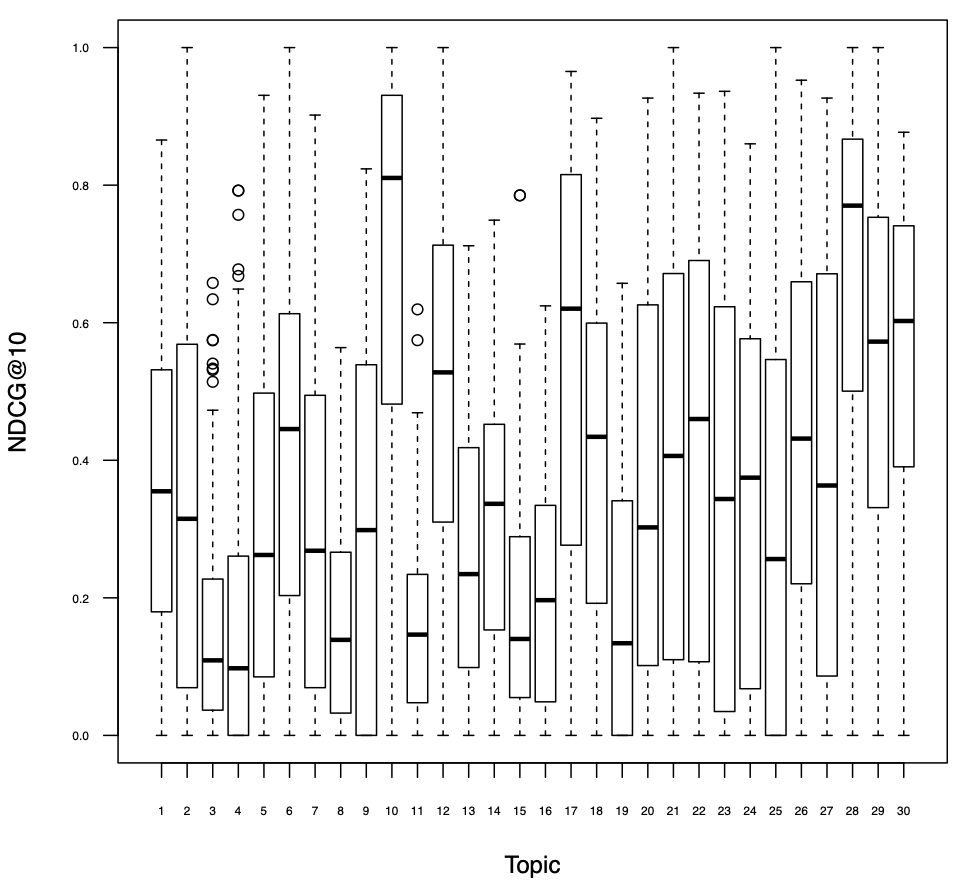}
\caption{Per-topic NDCG@10 scores aggregated over all Round 1 runs.}
\label{topicbox.fig}
\end{center}
\end{figure}

Figure~\ref{topicbox.fig} provides a view of how effective the participants' systems were as a group.
The figure contains a box-and-whiskers plot of the NDCG@10 scores across all 143 submitted runs for each topic.
The median NDCG@10 score for the topic is plotted as the black bar inside the box; the top and bottom of the box represent the third and first quartiles of the scores,respectively; the whiskers extend out another 1.5 times the interquartile range; and outliers are plotted as points.
Taken together, the set of systems was effective at finding relevant documents for every topic.   Eight topics have maximum NDCG@10 scores of 1.0, meaning at least one system retrieved ten fully relevant documents in the top 10 ranks.
Nonetheless, the spread of scores across runs is good for each topic
suggesting that the test collection is not too easy and that it
will be able to discriminate between systems.

\section{Challenges}
\label{sec:challenges}

The rapid organization of TREC-COVID and the shifting nature of knowledge of the underlying COVID-19 disease create challenges for organizing a TREC-style evaluation. In this section we list some of the limitations and cautions for using the TREC-COVID test collection.

The short time available for relevance judgments has had two main effects.
First, it puts a hard limit on the number of judgments that can be obtained, which when coupled with many participants results in quite shallow judgment pools.
Shallow pools in turn lead to incomplete judgment sets and relatively large uncertainty in systems' evaluation scores.
The measures for which the uncertainty is small, such as P@5, are not necessarily the best measures to understand the target user task.
Second, it requires using multiple assessors per topic.
This can damage the internal consistency of the judgments for a topic, since different assessors are likely to have different notions of relevance.

Capturing the dynamic nature of COVID-19 literature is a primary goal of TREC-COVID, and using a dynamic collection as the basis for a retrieval test collection presents its own challenges.
Changes come in different forms.
The papers themselves can change over time as preprints are updated, then published, or (potentially) retracted.  Papers may have only the abstract available at first, with the full text released sometime later. For example, of the 51K papers in the April 10 release of CORD-19 used in Round 1, 1041 papers had some change in the title, abstract, or full text by the May 1 release.  The vast majority of these changes were minor such as simple text edits. But 284 papers that had either no abstract or full text in the April 10 release did have those components in the May 1 release.  Such changes to a paper can invalidate previously made relevance judgments, so assessors re-judge changed papers for all previously-judged topics.  The relevance judgments in the final test collection will therefore not be a single judgment per document per topic as is the norm, but rather a series of judgments where each entry in the series is tied to a particular CORD-19 release through the round number.

An assessor's notion of what it means to be relevant may also change as the science around COVID-19 evolves.
The very public nature of the disease means that the assessment process cannot happen in a vacuum: assessors are aware of many of the latest medical developments, and that may alter their perception of relevance.
This impact is difficult to quantify empirically, though we do have anecdotal evidence from assessors regarding how they may have judged prior results differently based on what they now know.
This sort of re-assessment of prior judgments is {\em not} done.
The judgments need to be viewed as occurring in the time frame of its associated the CORD-19 release.

\section{Future}

 At the time of writing, TREC-COVID is entering Round 2 with five new topics, an updated document collection and the original set of judges. The immediate plans include three more evaluation rounds, which will bring us to Round 5 in July. By that time, we anticipate to better understand the trends in the growth of the document collection, retrieval of residual relevant documents, stability of the systems' ranking, and the interest of research community in continuing the evaluation. These factors will inform the organizers if additional rounds will be needed to build a permanent COVID-19 pandemic collection and resolve any remaining pandemic response system-building and evaluation questions. The completion of the last round and the analysis of the results will accomplish the first goal of TREC-COVID -- an evaluation of the systems and approaches that directly address information needs brought about by the COVID-19 pandemic. 

The final collection built over the entire set of the evaluation rounds will support the second goal of TREC-COVID: to discover approaches to managing scientific information and satisfy information needs of scientists during global outbreaks and similar health threats. The final collection will consist of the clearly marked versions of the data, topics and relevance judgments corresponding to each evaluation round. The cumulative relevance judgments in combination with the final document collection will allow researchers to develop approaches in a traditional post hoc use of TREC collections, probably with some caveats mentioned in section \ref{sec:challenges}. Whereas using the versions of the rounds as they evolve, the researchers can study approaches to addressing information needs as they evolve during pandemics. 

\bibliographystyle{plain}
\bibliography{ms}

\end{document}